\documentclass[aps,pre,graphicx,reprint,superscriptaddress]{revtex4-1}

\usepackage{amsmath}
\usepackage{amsfonts}
\usepackage{graphicx}
\usepackage{amssymb}
\usepackage{bm}
\usepackage{braket}
\usepackage{hyperref}
\usepackage{xcolor}

\begin{document}

\title{Formation of solitary zonal structures via the modulational instability of drift waves}

\author{Yao Zhou}
\email[]{yaozhou@princeton.edu}
\affiliation{Princeton Plasma Physics Laboratory, Princeton, New Jersey 08543, USA}

\author{Hongxuan Zhu}
\affiliation{Princeton Plasma Physics Laboratory, Princeton, New Jersey 08543, USA}
\affiliation{Department of Astrophysical Sciences, Princeton University, Princeton, NJ 08544, USA}

\author{I.~Y.~Dodin}
\affiliation{Princeton Plasma Physics Laboratory, Princeton, New Jersey 08543, USA}
\affiliation{Department of Astrophysical Sciences, Princeton University, Princeton, NJ 08544, USA}

\date{\today}

\begin{abstract}
The dynamics of the radial envelope of a weak coherent drift wave is approximately governed by a nonlinear Schr\"odinger equation, which emerges as a limit of the modified Hasegawa--Mima equation. The nonlinear Schr\"odinger equation has well-known soliton solutions, and its modulational instability can naturally generate solitary structures. In this paper, we demonstrate that this simple model can adequately describe the formation of solitary zonal structures in the modified Hasegawa--Mima equation, but only when the amplitude of the coherent drift wave is relatively small. At larger amplitudes, the modulational instability produces stationary zonal structures instead. Furthermore, we find that incoherent drift waves with beam-like spectra can also be modulationally unstable to the formation of solitary or stationary zonal structures, depending on the beam intensity. Notably, we show that these drift waves can be modeled as quantumlike particles (``driftons'') within a recently developed phase-space (Wigner--Moyal) formulation, which intuitively depicts the solitary zonal structures as quasi-monochromatic drifton condensates. Quantumlike effects, such as diffraction, are essential to these condensates; hence, the latter cannot be described by wave-kinetic models that are based on the ray approximation.

\end{abstract}

\maketitle

\section{Introduction}
In magnetically confined plasmas, radially propagating coherent structures are of great interest, as they can induce transport that is {ballistic} rather than {diffusive}. Examples include turbulence spreading \cite{Garbet1994,Lin2004,Hahm2004,Naulin2005,Guo2009} and avalanching \cite{Politzer2000,Candy2003,McMillan2009}, as well as the density ``blobs" in edge plasmas \cite{Zweben2007,Krasheninnikov2008,Krasheninnikov2016,Zhang2017}. Recently, such structures have also been identified in gyrokinetic simulations of subcritical plasmas \cite{VanWyk2016,VanWyk2017,McMillan2018}.

An arguably simplest model of radially propagating coherent structures considers the radial envelope dynamics of a weak {coherent} drift wave (DW) without forcing and dissipation. It has been shown that the envelope approximately follows a nonlinear Schr\"odinger equation (NLSE) \cite{Champeaux2001,Dewar2007}, where the (cubic) nonlinearity originates from the quasilinear interaction between the primary DW and a secondary zonal flow (ZF). The well-known soliton solution to the NLSE corresponds to a zonal structure that propagates radially at the DW group velocity. In particular, Guo et al.\,\cite{Guo2009} first studied this DW--ZF soliton in the context of turbulence spreading from linearly unstable regions to stable regions. Accordingly, they examined soliton formation due to the inhomogeneity of the linear growth rate.
However, an intrinsic mechanism for the formation of solitary zonal structures is still needed, since events such as turbulence avalanching seem to be dominated by local physics \cite{McMillan2009}. One natural candidate is the modulational instability (MI) \cite{Champeaux2001,Dewar2007}, which is known to generate solitary structures in the NLSE. [Here, ``solitary'' means propagating at a (roughly) constant speed while maintaining a (roughly) constant shape.] Nevertheless, the relevance of this mechanism in the modified Hasegawa--Mima equation (mHME) \cite{Dewar2007,Chandre2014}, the parent model of the NLSE, has remained unexplored.

In this paper, we explicitly demonstrate that the NLSE can adequately describe the formation of solitary zonal structures via the MI in the mHME, using both quasilinear and nonlinear simulations of the latter. However, these structures only emerge from primary DWs with relatively small amplitudes; at larger amplitudes, the MI produces stationary zonal structures instead. Then, as a generalization, we simulate the MI of incoherent DWs with beam-like spectra, using a recently developed Wigner--Moyal (WM) model of DW--ZF dynamics \cite{Ruiz2016}. While the finite beam width has a stabilizing effect on the MI, the results are similar to those from coherent DWs. That is, with moderate beam intensity, solitary zonal structures are formed; as the intensity increases, the zonal structures become stationary. 

One novelty of the WM model is that it treats DWs as quantumlike particles (``driftons'') and facilitates analysis of the zonal structures from an instructive phase-space perspective. The Wigner function of the DW--ZF solitons show concentration of DW quanta in both position and momentum, depicting them as self-trapping drifton condensates, akin to Bose--Einstein condensates in quantum mechanics. The MI-induced solitary zonal structures exhibit similar features, which suggests that they are essentially the DW--ZF solitons. In turn, it also implies that these structures are not the same as those obtained from wave-kinetic models \cite{Smolyakov1999,Trines2005}, which are the conventional tools for treating incoherent DWs. The reason is that diffraction, among other quantumlike effects, is fundamental to these condensates (in fact, it determines their sizes), but neglected by the ray approximation that the wave-kinetic models invoke \cite{Diamond2005}.

This paper is organized as follows. In Sec.\,\ref{sec:model}, we introduce the mHME and its quasilinear approximation. In Sec.\,\ref{sec:MI}, we consider coherent DWs by deriving the NLSE and its dispersion relation of the MI. In Sec.\,\ref{sec:ZI}, we describe the WM model and the MI of general DW spectra. Most of our new results are presented in Sec.\,\ref{sec:soliton}, where we study solitary zonal structures and their formation from coherent DWs and incoherent DW spectra. Our results are summarized and discussed in Sec.\,\ref{sec:summary}.

\section{Basic model}\label{sec:model}
\subsection{Modified Hasegawa--Mima equation}
In this paper, we study DWs within the modified Hasegawa--Mima equation (mHME) \cite{Dewar2007,Chandre2014}, which is the simplest yet useful model that captures many basic effects of interest. (The mHME has been referred to as the generalized \cite{Krommes2000} or extended \cite{Connaughton2015} Hasegawa--Mima equation as well.) In a dimensionless form, the mHME can be written as
\begin{subequations}\label{mHME}
\begin{align}
\partial_t w + \mathbf{v}\cdot\nabla w &- \beta\partial_y\varphi =0,\label{HME}\\
  w \doteq \nabla^2\varphi &-\tilde{\varphi}.\label{vorticity}
\end{align}
\end{subequations}
It is a 2D model in slab geometry, with coordinates $\mathbf{x}\equiv(x,y)$ and a uniform magnetic field normal to the plane. The gradient of the plasma density $n_0$ is in the radial ($x$) direction, and is parameterized by a (positive) constant $\beta\doteq a/L_{n}$, where $a$ is the system scale length and $L_{n}\doteq (- \mathrm{d}\ln n_0/\mathrm{d}x)^{-1}$ is the local scale length of the gradient. (The symbol $\doteq$ denotes definitions.) The ZF is in the poloidal ($y$) direction. Time $t$ is normalized by the transit time $a/c_\text{s}$, where $c_\text{s}$ is the sound speed. Space is normalized by the ion sound radius $\rho_\text{s}\doteq c_\text{s}/\Omega_\text{ci}$, where $\Omega_\text{ci}$ is the ion gyro-frequency. The electrostatic potential $\varphi(t,\mathbf{x})$ is normalized by $T_\text{e}\rho_\text{s}/(ea)$, where $e$ is the unit charge and $T_\text{e}$ is the electron temperature.
Accordingly, $\mathbf{v}\doteq(-\partial_y\varphi,\partial_x\varphi)$ is the $\mathbf{E}\times\mathbf{B}$ velocity.

In the mHME, the definition of the generalized vorticity $w$ \eqref{vorticity} involves separating the total $\varphi$ into the zonal component $\langle{\varphi}\rangle$ and non-zonal component $\tilde{\varphi}$. The former is the ``zonal average'' of $\varphi$, $\langle{\varphi}\rangle\doteq\int\mathrm{d}y\,\varphi/L_y$ (where $L_y$ is the domain length in $y$), and corresponds to the ZF. The latter is the fluctuating component, $\tilde{\varphi}\doteq\varphi-\langle{\varphi}\rangle$, and corresponds to DWs. The same notations apply to $w$ and $\mathbf{v}$ as well.

In contrast, in the original Hasegawa--Mima equation (oHME) \cite{Hasegawa1978}, the generalized vorticity is defined as $w \doteq \nabla^2\varphi-{\varphi}$. The modification in the mHME is due to the finding that the zonal potential $\langle{\varphi}\rangle$ does not contribute to the adiabatic electron response \cite{Dorland1993,Hammett1993}. The oHME is also called the Charney--Hasegawa--Mima equation for its equivalence to the Charney equation \cite{Charney1971}. Meanwhile, with $w \doteq \nabla^2\varphi$, Eq.\,\eqref{HME} becomes equivalent to the barotropic vorticity equation \cite{Charney1950}. Both the Charney equation and the barotropic vorticity equation are widely used in studies of geophysical fluids. The similarity between the mHME and these equations suggests that our results can, to an extent, be relevant to Rossby-wave turbulence in geophysics.

The mHME does not have a primary instability, i.e., an instability that generates DWs. Thus, external forcing is sometimes introduced as a proxy, and \textit{ad~hoc} dissipation must also be added to balance the energy input. 
However, due to the existence of the drift term (the linear term proportional to $\beta$), the mHME can support finite-amplitude DWs even in the absence of forcing. Then, ZFs can emerge from these DWs through a secondary instability, which is also known as the MI \cite{Champeaux2001,Dewar2007} or the zonostrophic instability \cite{Srinivasan2012}. In this paper, we focus on this particular process and will not be concerned with the origin of the DWs. Instead, we will introduce finite-amplitude DWs via initial conditions, and exclude forcing and dissipation (except briefly in the end of Sec.\,\ref{sec:FZI}).

\subsection{Quasilinear approximation}\label{sec:QL}
To proceed, let us separate Eq.\,\eqref{HME} into the non-zonal and zonal components, respectively:
\begin{subequations}\label{NL}
\begin{align}
\partial_t\tilde{w}+\langle\mathbf{v}\rangle\cdot\nabla \tilde{w}+ \tilde{\mathbf{v}}\cdot\nabla \langle{w}\rangle-\beta\partial_y\tilde{\varphi}&=\langle\tilde{\mathbf{v}}\cdot\nabla \tilde{w}\rangle-\tilde{\mathbf{v}}\cdot\nabla \tilde{w},\label{NLNZ}\\
\partial_t \langle{w}\rangle + \langle\tilde{\mathbf{v}}\cdot\nabla \tilde{w}\rangle&=0.\label{NLZ}
\end{align}
\end{subequations}
In the studies of ZF formation, it is common to assume that the nonlinearity on the right-hand side (RHS) of Eq.\,\eqref{NLNZ} is weak, which can be physically interpreted as neglecting the eddy--eddy interactions between the DWs. This is called the quasilinear approximation, for it leads to two linear equations that are nonlinearly coupled:
\begin{subequations}\label{QL}
\begin{align}
\partial_t\tilde{w}+U\partial_y\tilde{w}-(\beta+U'')\partial_y\tilde{\varphi}&=0,\label{quasilinearNZ}\\
\partial_t U -\partial_x\langle{\partial_x\tilde{\varphi}  \partial_y\tilde{\varphi}}\rangle&=0.\label{quasilinearZ}
\end{align}
\end{subequations}
For convenience, we introduce the ZF velocity $U(t,x)\doteq \langle v_y\rangle =  \partial_x\langle\varphi\rangle$, with $U''\doteq\partial^2_xU$.  

The quasilinear mHME \eqref{QL} has been shown to reproduce many of the basic features of the original nonlinear system \eqref{NL}, at least qualitatively \cite{Srinivasan2012}. Hence, we consider the quasilinear approximation sufficient for our purposes, and adopt it throughout the rest of the paper (except in Figs.\,\ref{delta} and \ref{MI}, where we briefly present some nonlinear simulation results as verifications for the corresponding quasilinear simulations).

The non-zonal component of the quasilinear mHME \eqref{quasilinearNZ} can also be written in the form of a Schr\"odinger equation for DW quanta (driftons),
\begin{align}
i\partial_t\tilde{w}=\hat{H}\tilde{w}.\label{LSE}
\end{align}
Unlike the truly quantum Schr\"odinger equation, Eq.\,\eqref{LSE} does not contain $\hbar$, so it is purely classical. (Likewise, $|\tilde{w}|^2$ is an action density rather than number density.) Also, the Hamiltonian operator, 
\begin{align}
\hat{H}\doteq(\beta+\hat{U}'') \hat{p}_y\hat{\bar{p}}^{-2} + \hat{U}\hat{p}_y,\label{Hoperator}
\end{align}
is not entirely Hermitian, since $\hat{U}''$ does not commute with $\hat{\bar{p}}^{-2}$. Here, $\hat{\mathbf{p}}\doteq-i\nabla$ can be understood as the momentum (wave-vector) operator, and $\hat{\bar{p}}^{2}\doteq1+\hat{p}_x^{2}+\hat{p}_y^{2}$ such that $\tilde{w}=-\hat{\bar{p}}^{2}\tilde{\varphi}$. Also, $\hat{U}\doteq U(t,\hat{x})$ and $\hat{U}''\doteq U''(t,\hat{x})$, with $\hat{\mathbf{x}}$ being the position operator. This quantumlike formalism proves useful in deriving the nonlinear Schr\"odinger equation that governs the envelope dynamics of coherent DWs (Sec.\,\ref{sec:MI}), as well as the WM formulation that can describe incoherent DWs (Sec.\,\ref{sec:ZI}). 

\section{Coherent drift waves}\label{sec:MI}
The quasilinear mHME \eqref{QL} has an exact plane-wave solution with finite amplitude, $\tilde{w}=\text{Re}(\psi_0e^{i\mathbf{k}\cdot\mathbf{x}-i\Omega t})$, where $\mathbf{k}\doteq(k_x,k_y)$ is the wave-vector and $\Omega\doteq\beta k_y/\bar{k}^2$ is the DW frequency, with $\bar{k}^2\doteq 1+k_x^2+k_y^2$ and $\psi_0$ being a complex constant denoting the amplitude.

This primary wave, when subject to large-scale modulations, can become unstable. One simplified way to study this MI is to consider the envelope dynamics of a coherent DW. The first study of such kind appears to be Ref.\,\cite{Nozaki1979}, which is based on the oHME and only considers poloidal modulations (``streamers''). In Ref.\,\cite{Champeaux2001}, the mHME is employed, and radial and poloidal modulations are treated on the same footing. More comprehensive reviews of the envelope formalism can be found in Refs.\,\cite{Diamond2005,Dewar2007}. Also notably, related equations were later rediscovered independently in Refs.\,\cite{Guo2009} (with over-simplified coefficients) and \cite{Jovanovic2010}.

In all of these studies, it is noticed that the envelope equation can be approximated as a NLSE, which is well-known to have a MI. In Sec.\,\ref{sec:NLSE}, we show how the NLSE follows naturally from our quantumlike formalism. In Sec.\,\ref{sec:NLSED}, we rederive the corresponding dispersion relation of the MI.

\subsection{Nonlinear Schr\"odinger equation}\label{sec:NLSE}

Let us represent the Hamiltonian operator \eqref{Hoperator} as $\hat{H} = \hat{H}_0 + \hat{H}_U$, where $\hat{H}_0$ is the $U$-independent part and $\hat{H}_U$ scales linearly with $U$. Since we focus on zonal structures, here we consider a coherent DW with radial modulation only, $\tilde{w}=\text{Re}[\psi(t,x)e^{i\mathbf{k}\cdot\mathbf{x}-i\Omega t}]$. We assume that the envelope $\psi$ is slow, i.e., $|\partial_x\ln\psi|\ll |k_x|$, and also that $|U|$ is small. Then, the Hamiltonian operator \eqref{Hoperator} can be approximated as
\begin{equation}
\hat{H}_0\approx\Omega+v_\text{g}\hat{p}_x+(\chi/2)\hat{p}_x^2,\quad\hat{H}_U \approx k_yU,\label{Happrox}
\end{equation}
where $v_\text{g}\doteq\partial\Omega/\partial k_x$ is the radial group velocity and $\chi\doteq\partial^2\Omega/\partial k_x^2$, or explicitly,
\begin{equation} 
v_\text{g}= -2\beta k_xk_y/\bar{k}^4,\quad \chi= (2\beta k_y/\bar{k}^6)(4k_x^2-\bar{k}^2).\label{vgchi}
\end{equation}
Equation \eqref{Happrox} can be viewed as a special case of the Weyl expansion derived in Ref.\,\cite{Dodin2019} for an inhomogeneous medium. Additionally, the term proportional to $U''$ has been neglected because both $|U|$ and $\partial^2_x$ are assumed small. Then, the resulting equation for $\psi$ is
\begin{align}
i(\partial_t  +v_\text{g} \partial_x)\psi \approx -(\chi/2)\partial_x^2\psi+k_yU \psi.\label{LSEapprox}
\end{align}

Meanwhile, using $\tilde\varphi\approx-\tilde w/\bar{k}^2$, we can approximate the ZF equation \eqref{quasilinearZ} as
\begin{align}
\partial_t U \approx \partial_x\langle\partial_x\tilde w\,\partial_y\tilde w\rangle/\bar{k}^4 \approx k_xk_y\partial_x|\psi|^2/(2\bar{k}^4),\label{ZFapprox}
\end{align}
where the factor $1/2$ originates from zonal averaging. From Eq.\,\eqref{LSEapprox}, we can see that, to the leading order, the modulation propagates at the group velocity $v_\text{g}$. Hence, we can assume that $\partial_t \approx -v_\text{g} \partial_x$ in Eq.\,\eqref{ZFapprox}, and integrate in $x$ to obtain $U \approx-k_xk_y|\psi|^2/(2\bar{k}^4v_\text{g})$, or more explicitly,
\begin{align}
U \approx|\psi|^2/(4\beta). \label{EOS}
\end{align}
Here, vanishing boundary condition in $x$ is implied. For other boundary conditions (e.g., $\langle\varphi\rangle$ periodic in $x$), an additional integration constant may be needed on the RHS of Eq.\,\eqref{EOS} for consistency. (This constant can be easily removed by a Galilean transformation, however.) Also, with $|U|$ assumed small, it is implied that the DW amplitude $|\psi|$ should be small too. Substituting Eq.\,\eqref{EOS} into Eq.\,\eqref{LSEapprox}, we obtain
\begin{align}
i(\partial_t +v_\text{g} \partial_x)\psi \approx -(\chi/2)\partial_x^2\psi+k_y|\psi|^2 \psi/(4\beta).\label{NLSE}
\end{align}

Equation \eqref{NLSE} has previously been derived (using somewhat different approaches) in Refs.\,\cite{Champeaux2001,Dewar2007}. It has the form of a NLSE, or the Gross--Pitaevskii equation, so the structures it describes can be viewed as ``drifton condensates'' (by analogy with the Bose--Einstein condensate). Namely, Eq.\,\eqref{NLSE} shows that it is energetically favorable for driftons to be in a correlated state rather than have random phases. Also, Eq.\,\eqref{EOS} can be interpreted as an equation of state of the condensates, as it provides a local relation between the drifton density $|\psi|^2/2$ and another ``thermodynamic'' property of the condensates, $U$.

\begin{figure}
 \includegraphics[width=\columnwidth]{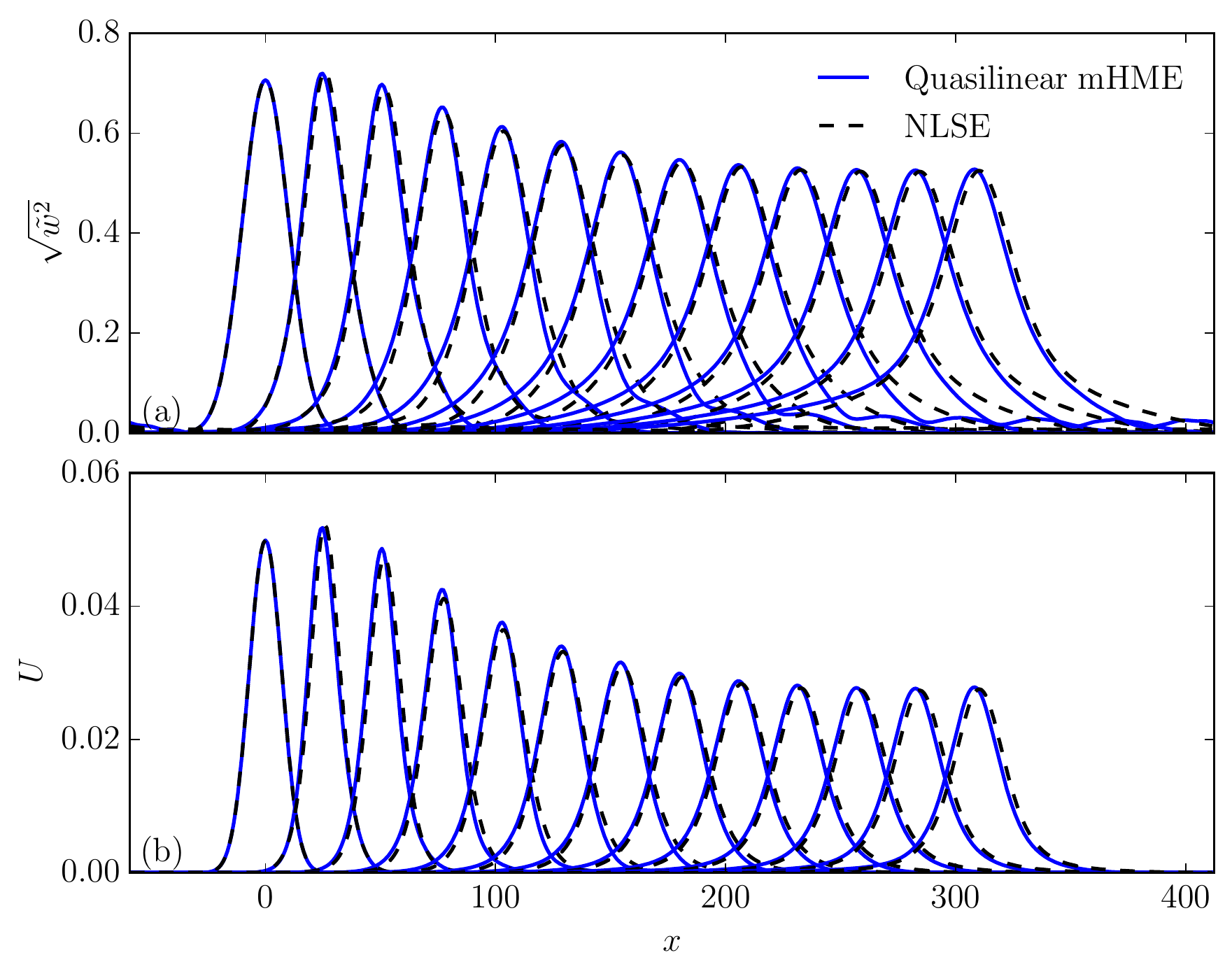}
 \caption{Sequences of (a) the DW envelope $\sqrt{\tilde{w}^2}$ ($|\psi|/\sqrt{2}$) and (b) the ZF velocity $U$ [$|\psi|^2/(4\beta)$] obtained from NLSE (dashed) and quasilinear mHME (solid) simulations. From left to right, the snapshots are taken at $t=0, 40, \ldots, 480$, respectively. The initial condition is a Gaussian envelope $\psi=2\eta\sqrt{-{\beta\chi}/{k_y}}{\exp(-\eta^2 x^2/2)}$ with $\eta=0.1$. We use the following parameters here and in all other figures throughout the paper: $\beta=5$, $k_x=0.3$, and $k_y=-0.3$.}\label{envelope}
\end{figure}

In Fig.\,\ref{envelope}, we compare the evolution of an initially Gaussian envelope in numerical simulations of the NLSE and the quasilinear mHME. The good agreement between the solutions confirms that the former is a reasonable approximation of the latter. All of our simulations using configuration-space models (the NLSE, the quasilinear mHME, and the nonlinear mHME) are pseudo-spectral, dealiased, and performed on periodic domains.

\subsection{Modulational instability}\label{sec:NLSED}
The NLSE \eqref{NLSE} has an exact homogeneous solution $\psi = \psi_0\exp[-ik_y|\psi_0|^2t/(4\beta)]$. The frequency $\omega$ and wavenumber $q$ of a linear perturbation on this solution satisfies the following dispersion relation \cite{Dewar2007},
\begin{align}
(\omega-qv_\text{g})^2=\frac{\chi^2 q^4}{4}\left(1+\frac{k_y|\psi_0|^2}{\beta\chi q^2}\right) .\label{NLSED}
\end{align}
When $\beta\chi/k_y\propto 4k_x^2-\bar{k}^2<0$, the frequency is complex for small $q$ and the solution is linearly unstable, with the wavenumber of the fastest-growing mode given by $q_\text{max} = |\psi_0|\sqrt{-k_y/(2\beta\chi)}$. This is the well-known MI of the NLSE, arising from the interplay of diffraction and self-focusing [the first and second terms on the RHS of Eq.\,\eqref{NLSE}, respectively].

The NLSE \eqref{NLSE} offers an intuitive perspective on the MI of coherent DWs. However, as an approximate model, it is restricted to slow modulations and small $|\psi_0|$. When $|\psi_0|$ is large, $q_\text{max}$ can be comparable or larger than $k_x$, which is inconsistent with the underlying assumption of the NLSE. In addition, the NLSE only applies to primary waves with non-zero radial group velocity $v_\text{g}\propto k_xk_y$. While in this paper we focus on such waves for this very feature, primary waves with $k_x=0$ are also of interest, since they correspond to the fastest growing modes in some primary instabilities, particularly, ion-tempatature-gradient modes \cite{Chen2000,Rogers2000}. 

In fact, there are more general approaches to deriving the dispersion relation of the MI. One way is to employ the four-mode truncation (4MT) method. As the name suggests, the 4MT is a truncation of the mHME in spectral representation by only keeping four modes: a primary wave with wave-vector $\mathbf{k}$, a modulation with wave-vector $\mathbf{q}$, and two sidebands with wave-vectors $\mathbf{k}\pm\mathbf{q}$. In general, the modulation does not have to be purely radial. For example, the MI with a purely radial primary wave (a ZF) and a purely poloidal modulation is a tertiary instability of the ZF \cite{Zhu2018,Zhu2018a}. Detailed discussions on the 4MT can be found in Refs.\,\cite{Connaughton2010,Gallagher2012,Zhu2019}. Meanwhile, for purely radial modulations with $\mathbf{q}=(q,0)$, which we focus on in this paper, the MI of coherent DWs can be considered as a special case of the MI of general DW spectra, which is discussed in Sec.\,\ref{sec:ZI}.

\section{Drift-wave ensembles}\label{sec:ZI}
Equation \eqref{LSE}, along with Eq.\,\eqref{quasilinearZ}, governs the quasilinear dynamics of a single realization of DWs (in quantum mechanical terms, a ``pure state''). However, due to the incoherent nature of DW turbulence, it is useful to consider the dynamics of an ensemble of DWs statistically. This is equivalent to studying the von Neumann equation that follows from the Schr\"odinger equation \eqref{LSE}, which can describe the dynamics of ``mixed states''. In double-configuration-space representation, this leads to the theory of second-order cumulant expansion (CE2), which has been widely used in geophysical fluids (e.g., \cite{Farrell2003,Marston2008,Srinivasan2012,Parker2014}) and subsequently introduced to plasma physics \cite{Farrell2009,Parker2013}.
A mathematically equivalent yet physically more intuitive alternative to the CE2 model is the phase-space representation of the von Neumann equation. This leads to the Wigner--Moyal (WM) model of DW--ZF dynamics, which we introduce in Sec.\,\ref{sec:WM}. The WM model can describe the MI of general DW spectra, which is presented in Sec.\,\ref{sec:ZID}.

\subsection{Wigner--Moyal formulation}\label{sec:WM}
The WM model of DW--ZF dynamics was first derived in Ref.\,\cite{Ruiz2016}. The derivation starts from the quasilinear mHME \eqref{QL} and leads to the following equations:
\begin{subequations}\label{WME}
\begin{align}
\partial_t W &= \{\!\{\mathcal{H},W\}\!\}+[[\Gamma,W]],\label{DW}\\
\partial_t U &= \partial_x\int\frac{\mathrm{d}^2p}{(2\pi)^2}\frac{1}{\bar{p}^{2}} \star p_x p_y W \star \frac{1}{\bar{p}^{2}}.\label{ZF}
\end{align}
\end{subequations}
Here, $\mathbf{p}\doteq(p_x,p_y)$ is the coordinate in the DW momentum (wave-vector) space, and $W(t,x,\mathbf{p})$ is the zonal-averaged Wigner function \cite{Wigner1932}. For a single realization of $\tilde{w}(t,\mathbf{x})$, $W$ can be written as
\begin{align}
W\doteq\left\langle\int\mathrm{d}^2s\,e^{-i\mathbf{p}\cdot\mathbf{s}}\tilde{w}\left(t,\mathbf{x}+\frac{\mathbf{s}}{2}\right)\tilde{w}\left(t,\mathbf{x}-\frac{\mathbf{s}}{2}\right)\right\rangle.\label{WF}
\end{align}
For an ensemble of realizations, the zonal average (again, denoted by the angle bracket) can be regarded as an ensemble average. The Wigner function $W$ is a quasi-probability distribution of driftons, and the ZF velocity $U$ serves as a collective field through which they interact.
Since $\tilde{w}$ is real (unlike in quantum mechanics where the wave functions are complex), the DW Wigner function has a unique parity in $\mathbf{p}$ that $W(t,x,\mathbf{p})=W(t,x,\mathbf{-p})$. This implies that driftons come in pairs, i.e., each drifton with wave-vector $\mathbf{p}$ has a twin with wave-vector $-\mathbf{p}$.

The specific dynamics of the driftons is governed by
\begin{subequations}\label{HG}
\begin{align}
\mathcal{H} &= \beta p_y/\bar{p}^2 + p_y U + [[U'',p_y/\bar{p}^2]]/2,\label{Hamiltonian}\\
\Gamma &= \{\!\{U'',p_y/\bar{p}^2\}\!\}/2,\label{Gamma}
\end{align}
\end{subequations}
which are the Hermitian and anti-Hermitian parts of the Hamiltonian, respectively. Here, the Moyal star product $A(\mathbf{x},\mathbf{p})\star B(\mathbf{x},\mathbf{p})\doteq A\exp(i\hat{\mathcal{L}}/2)B$ \cite{Moyal1949}, where the Janus operator $\hat{\mathcal{L}}$ is defined as $A\hat{\mathcal{L}} B\doteq (\partial_\mathbf{x}A)\cdot(\partial_\mathbf{p}B) - (\partial_\mathbf{p}A)\cdot(\partial_\mathbf{x}B)$.
The Moyal sine and cosine brackets are given by $\{\!\{A,B\}\!\}\doteq 2A\sin(\hat{\mathcal{L}}/2)B$ and $[[A,B]]\doteq 2A\cos(\hat{\mathcal{L}}/2)B$, respectively. A detailed derivation of Eq.\,\eqref{WME} and a review of the Weyl calculus can be found in Ref.\,\cite{Ruiz2016}. 

The WM equation \eqref{WME} can be understood as a kinetic model treating driftons as quantumlike particles with finite ``de~Broglie'' wavelengths. As such, it captures ``full-wave'' effects missing in wave-kinetic models of DW--ZF dynamics based on the ray approximation \cite{Smolyakov1999,Parker2016}, which treat driftons as point particles with zero wavelengths. While the wave-kinetic models prove useful in some scenarios \cite{Parker2018,Zhu2018b,Ruiz2019}, they are insufficient for the problems that we study in this paper. A detailed discussion on the limitations of the wave-kinetic models is presented in Appendix\,\ref{sec:WKE}.

\subsection{Modulational instability}\label{sec:ZID}

In the WM model, a statistically homogeneous equilibrium can be described by a Wigner function $W=\mathcal{W}(\mathbf{p})$, which can be interpreted as a DW spectrum. Linearizing Eq.\,\eqref{WME} about $\mathcal{W}(\mathbf{p})$ leads to the following dispersion relation of the MI \cite{Ruiz2016,Zhu2018}, 
\begin{align}
\omega =\sum_{\pm}\int\frac{\mathrm{d}^2p}{(2\pi)^2}\mathcal{W}(\mathbf{p})\frac{\mp qp_y^2(p_x\pm q/2)(1-{q^2}/{\bar{p}^2})}{ \bar{p}^2\bar{p}_{\pm }^2\omega+2\beta qp_y(p_x\pm q/2)},\label{ZID}
\end{align}
where $\bar{p}_{\pm }^2\doteq 1+(p_x\pm q)^2+p_y^2$. The MI of a monochromatic DW $\tilde{w}=\text{Re}(\psi_0e^{i\mathbf{k}\cdot\mathbf{x}-i\Omega t})$ can be considered as a special case. Using Eq.\,\eqref{WF}, we can obtain the corresponding spectrum of this DW
\begin{equation}
\mathcal{W}(\mathbf{p})=\pi^2|\psi_0|^2[\delta(\mathbf{p}+\mathbf{k})+\delta(\mathbf{p}-\mathbf{k})].\label{2delta}
\end{equation}
Substituting Eq.\,\eqref{2delta} into Eq.\,\eqref{ZID} leads to the dispersion relation of the MI of this monochromatic DW,
\begin{align}
\prod_{\pm}&\left[ \bar{k}^2\bar{k}_{\pm }^2\omega+2\beta qk_y(k_x\pm q/2)\right]=\nonumber\\
&{|\psi_0|^2} q^2k_y^2({\bar{k}^2}-{q^2})(4k_x^2-{\bar{k}^2}-{q^2})/2.\label{MID}
\end{align}
Note that this dispersion relation agrees with that given by the 4MT method \cite{Connaughton2010,Gallagher2012}, and is the exact dispersion relation of the quasilinear mHME \eqref{QL}. 

\begin{figure}
 \includegraphics[width=\columnwidth]{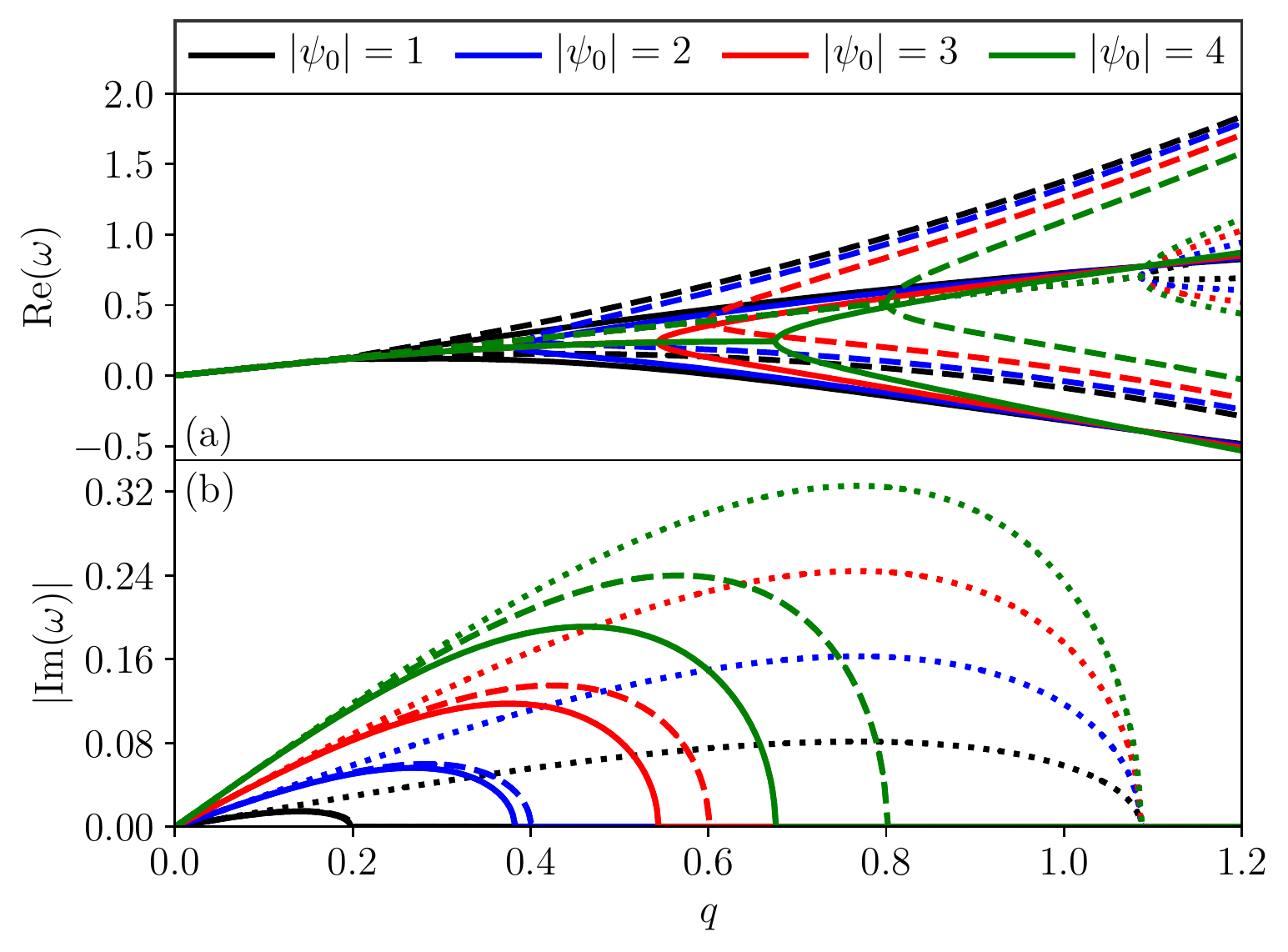}
 \caption{The real (a) and imaginary (b) parts of the dispersion relations of the MI of monochromatic DWs with various amplitudes ($|\psi_0|$, labeled by colors). Three different models are used: the WM model [solid, Eq.\,\eqref{MID}], the NLSE [dashed, Eq.\,\eqref{NLSED}], and a wave-kinetic model [dotted, Eq.\,\eqref{WKD}]. Since the complex solutions come as complex conjugates, we only show in (b) the absolute values of the imaginary parts, i.e., the growth rates. The same parameters as in Fig.\,\ref{envelope} are used.}\label{dispersion}
\end{figure}

It is instructive to compare the exact dispersion relation \eqref{MID} with the approximate one \eqref{NLSED} obtained from the NLSE. In Fig.\,\ref{dispersion}, we show both dispersion relations (for a given $\mathbf{k}$) with multiple values of $|\psi_0|$. It can be seen that the agreement between the dispersion relations is better for small $|\psi_0|$ and small $q$, which is consistent with the fact that the NLSE is derived based on the assumptions of slow modulation and small DW amplitude.
Still, in Fig.\,\ref{dispersion}, the NLSE appears to be a reasonable approximation of the quasilinear mHME even for relatively large $|\psi_0|$, in terms of capturing the linear MI. (This is not the case in the nonlinear stage of the MI, as we will discuss in Sec.\,\ref{sec:soliton}.) For comparison, Fig.\,\ref{dispersion} also shows the dispersion relations obtained from a wave-kinetic model based on the ray approximation (overviewed in Appendix\,\ref{sec:WKE}), which do not accurately approximate the exact ones. The reason for the discrepancy is that the wave-kinetic model misses essential full-wave (quantumlike) effects, particularly diffraction.

From Fig.\,\ref{dispersion}, we can see that the unstable modulations have finite real frequencies due to the DW group velocity, and hence propagate while growing. This feature has largely been overlooked in the past, possibly because zonal structures are typically perceived as (quasi-) stationary. Nevertheless, as we will show in Sec.\,\ref{sec:soliton}, the propagation of these zonal modes can have consequences in the nonlinear stage of the MI, leading to the formation of solitary zonal structures.

\section{Solitary zonal structures}\label{sec:soliton}

\subsection{Drift-wave--zonal-flow soliton}\label{sec:solution}

Since Eq.\,\eqref{NLSE} is of the same form as the well-known NLSE, it has the usual soliton solution

\begin{align}
\psi=2\eta\sqrt{-\frac{\beta\chi}{k_y}}\frac{\exp{(i\chi\eta^2t/2)}}{\cosh[\eta(x-v_\text{g}t)]}.\label{psiS}
\end{align}
Here, $\eta$ is a free parameter that can be regarded as the soliton inverse width.
With $\tilde{w}=\text{Re}[\psi(t,x)e^{i\mathbf{k}\cdot\mathbf{x}-i\Omega t}]$ and the equation of state \eqref{EOS},
Eq.\,\eqref{psiS} translates to an approximate soliton solution to the quasilinear mHME \eqref{QL}: 
\begin{subequations}\label{soliton}
\begin{align}
\tilde{w}&=2\eta\sqrt{-\frac{\beta\chi}{k_y}}\frac{\cos{[\mathbf{k}\cdot\mathbf{x}-(\Omega -\chi\eta^2/2)t]}}{\cosh[\eta(x-v_\text{g}t)]},\label{wS}\\
U &=\frac{-\eta^2\chi}{k_y\cosh^2[\eta(x-v_\text{g}t)]}. \label{US}
\end{align}
\end{subequations}
To our knowledge, this DW--ZF soliton was first explicitly discussed in Ref.\,\cite{Guo2009} and then in Ref.\,\cite{Jovanovic2010}, even though the NLSE \eqref{NLSE} that governs the DW envelope dynamics had been derived earlier \cite{Champeaux2001,Dewar2007}. It is fundamentally different from the vortex-pair solution called ``modon'' \cite{Meiss1983,Horton1994}, which is a 2D structure that propagates poloidally, whereas the DW--ZF soliton is an essentially 1D structure that propagates radially.

\begin{figure}
 \includegraphics[width=\columnwidth]{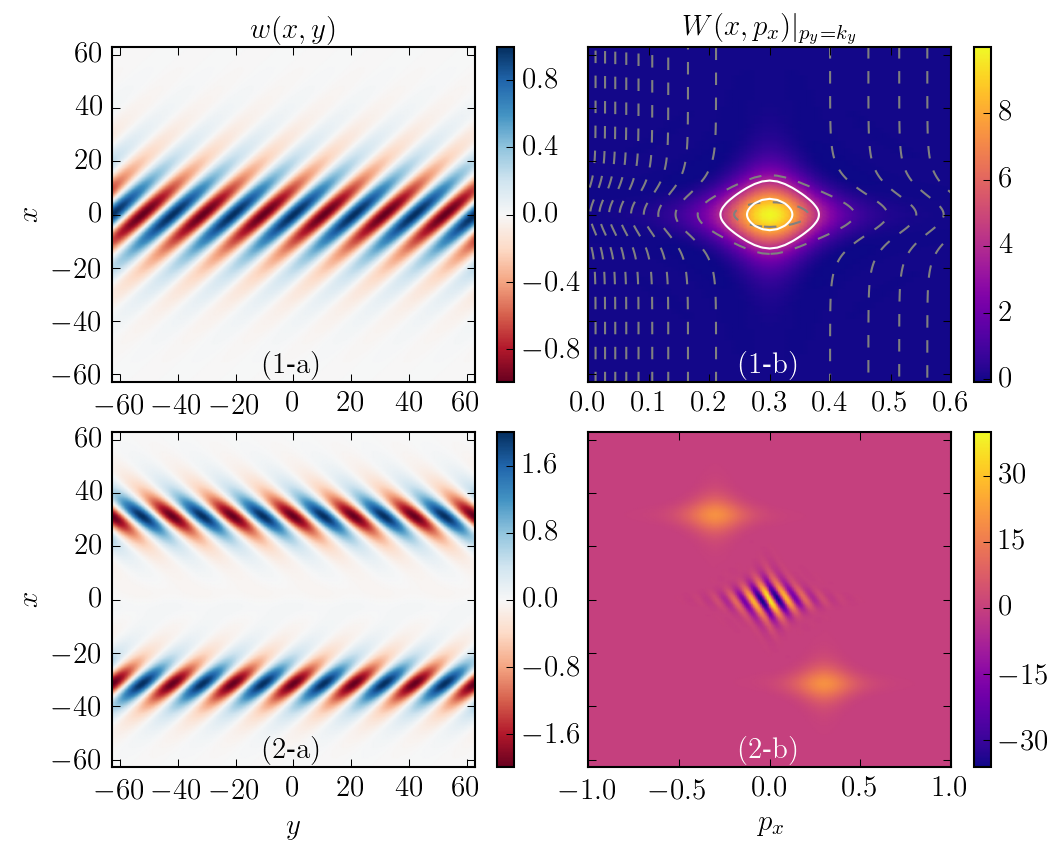}
 \caption{Snapshots of the DW--ZF soliton \eqref{soliton} with $\eta=0.1$ (row 1) and two solitons with $\eta=0.2$ and opposing $k_x$ (row 2). Column (a) shows the generalized vorticity $w(x,y)$, while column (b) shows the Wigner function $W(x,p_x)|_{p_y=k_y}$. The white solid contours in (1-b) are of the Wigner function, and the gray dashed ones are of the Hamiltonian in the moving frame, $\mathcal{H}_\text{m}$. The same parameters as in Fig.\,\ref{envelope} are used.
 }\label{solitonW}
\end{figure}

In Fig.\,\ref{solitonW}, we show snapshots of a DW--ZF soliton in both configuration space (1-a) and phase space (1-b). The Wigner function $W$, obtained using Eq.\,\eqref{WF}, is concentrated in both position and momentum, illustrating that the soliton is a quasi-monochromatic drifton condensate. It is also shown that the contours of $W$ do not coincide with those of the Hamiltonian in the moving frame $\mathcal{H}_\text{m}\doteq\mathcal{H}-v_\text{g}p_x$. This distinguishes the DW--ZF soliton from the BGK-type structures obtained from wave-kinetic models based on the ray approximation (Appendix \ref{sec:WKE}), where $W$ is a function of $\mathcal{H}_\text{m}$ only \cite{Smolyakov2000,Kaw2002,Itoh2005,Singh2014}. After all, the DW--ZF soliton is a result of the balance between self-focusing and diffraction, and the latter is a quantumlike effect missing in wave-kinetic models \cite{Diamond2005}. Therefore, wave-kinetic models cannot describe the DW--ZF soliton \eqref{soliton}, even though the requirement that its envelope be slow may seem consistent with the ray approximation.

Also shown in Fig.\,\ref{solitonW} (row 2) are snapshots of two superposing solitons with wave-vectors $(k_x, k_y)$ and $(-k_x, k_y)$, respectively. Accordingly, these solitons have opposing group velocities and hence counter-propagate. Upon colliding, they tunnel through each other. The striations between the drifton condensates in (2-b) are signatures of the quantum superposition of macroscopically distinct states, i.e., ``cat states'' \cite{Kenfack2004,Weinbub2018}.

\begin{figure*}
  \centering
 \includegraphics[width=\textwidth]{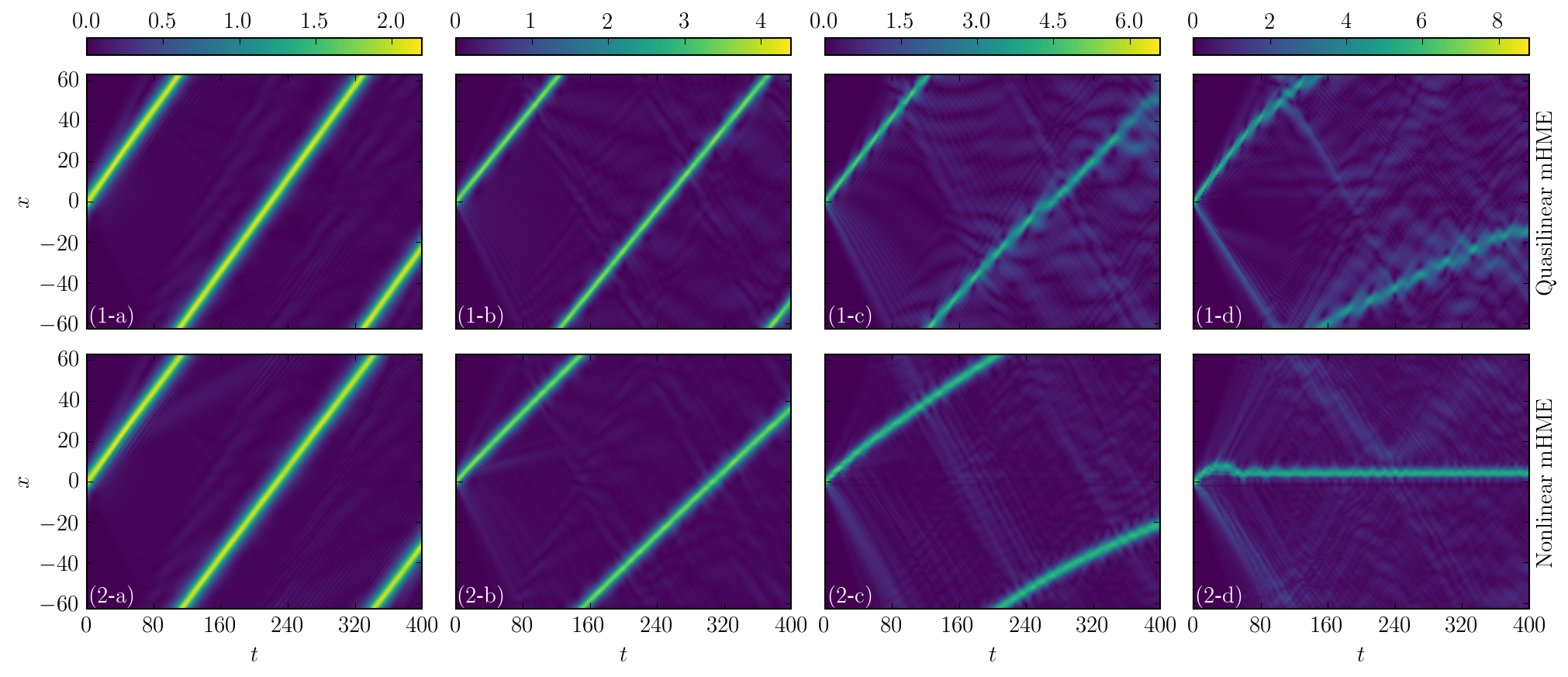}
 \caption{Quasilinear (row 1) and nonlinear (row 2) mHME simulations initialized with the soliton solution \eqref{soliton}. The columns correspond to various values of the inverse width: (a) $\eta= 0.3$, (b) $\eta= 0.6$, (c) $\eta= 0.9$, and (d) $\eta= 1.2$. The colormaps show the spatial-temporal evolution of the DW envelope $\sqrt{\langle\tilde{w}^2\rangle}$. The same parameters as in Fig.\,\ref{envelope} are used.
 }\label{delta}
\end{figure*}

In principle, $\eta\ll |k_x|$ is required for the soliton solution \eqref{soliton} to stand in the quasilinear mHME, since the NLSE is derived under the assumptions of slow envelope modulation and small DW amplitude. In practice, however, we find that the solitary behavior of this solution is quite robust, and extends even to $\eta\sim |k_x|$. In Fig.\,\ref{delta} (row 1), we show the spatial-temporal evolution of the DW envelope from quasilinear mHME simulations initialized with Eq.\,\eqref{soliton} for various values of $\eta$. As $\eta$ is doubled from (1-a) to (1-b), the zonal structure keeps propagating much like a soliton, while some small-amplitude structures emerge and the speed decreases slightly. As $\eta$ is tripled and quadrupled in (1-c) and (1-d), respectively, the zonal structure gradually breaks down and eventually stops propagating. 

In Fig.\,\ref{delta} (row 2), we also show corresponding results from nonlinear mHME simulations to verify the quasilinear simulations. The same qualitative features can be observed: the solitary behavior is robust at relatively small $\eta$, while the propagation eventually stops as $\eta$ keeps increasing. Admittedly, at larger $\eta$, the propagation stops more quickly in the nonlinear simulations. Nonetheless, for solitary structures with smaller $\eta$, which we focus on in this paper, the quasilinear approximation proves reasonable and acceptable. 

Having confirmed that the mHME indeed allows for solitary zonal structures, next, we demonstrate how they can spontaneously form via the MI, from either coherent DWs (Sec.\,\ref{sec:FMI}) or incoherent DW spectra (Sec\,\ref{sec:FZI}). These are the main results of this paper. 

\subsection{Zonal structures from coherent drift waves}\label{sec:FMI}
\begin{figure*}
 \includegraphics[width=\textwidth]{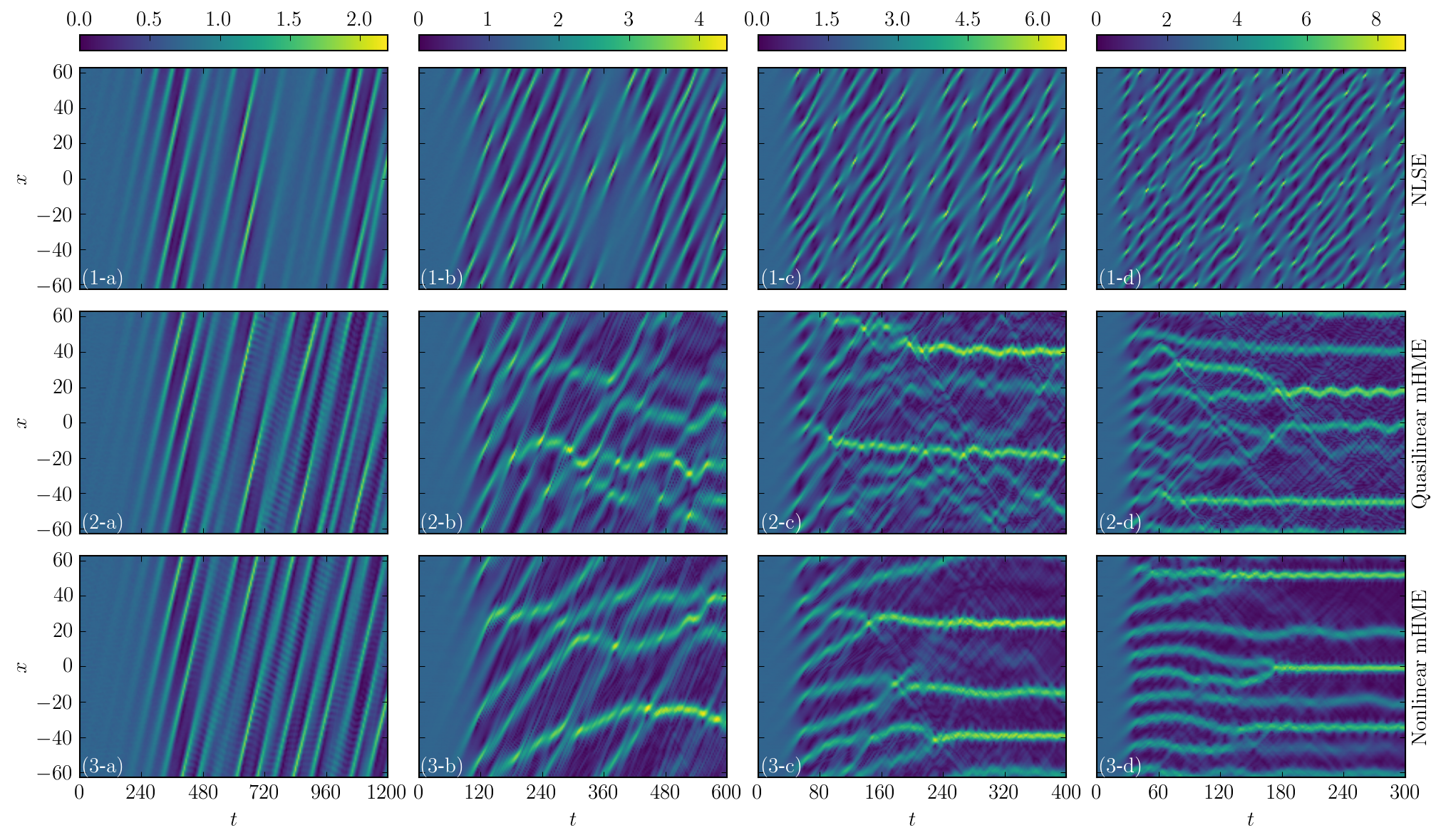}
 \caption{Simulations of the MI using different models: the NLSE (row 1), the quasilinear mHME (row 2), and the nonlinear mHME (row 3). The columns correspond to various values of the primary-wave amplitude: (a) $|\psi_0|= 1$, (b) $|\psi_0|= 2$, (c) $|\psi_0|= 3$, and (d) $|\psi_0|= 4$. The colormaps show the spatial-temporal evolution of the DW envelope $\sqrt{\langle\tilde{w}^2\rangle}$ ($|\psi|/\sqrt{2}$ in the NLSE). The simulations are run for different total time due to the differences in the growth rates. The same parameters as in Fig.\,\ref{envelope} are used. }\label{MI}
\end{figure*}

In Fig.\,\ref{MI}, we present numerical simulations of the MI, through the nonlinear stage, using three different models: the NLSE, the quasilinear mHME, and the nonlinear mHME. The initial states are chosen to be primary waves with various amplitudes ($|\psi_0|$) and random perturbations on top to seed the instability. The amplitudes of the perturbations are the same in all simulations.

In the NLSE simulations (row 1), for all values of $|\psi_0|$, the MI leads to the formation of zonal structures that behave
approximately as solitons, notwithstanding the increasingly apparent nonlinear oscillations and interactions with increasing $|\psi_0|$. The wavenumbers of the structures are consistent with the fastest-growing wavenumbers in Fig.\,\ref{dispersion}. The amplitudes of the structures increase with $|\psi_0|$ while the widths decrease, which is also consistent with the properties of the soliton solution \eqref{psiS}. Note that in (1-c) and (1-d), the amplitudes of the structures are already comparable to those that stop propagating in Figs.\,\ref{delta}(c) and \ref{delta}(d), respectively. The implication is that in the corresponding mHME simulations, we may not be able to observe these solitary zonal structures. 

Indeed, this can be seen in both the quasilinear (row 2) and nonlinear (row 3) simulations in Fig.\,\ref{MI}. At small $|\psi_0|$ [column (a)], the spatial-temporal evolutions of the DW envelope closely resemble their NLSE counterpart (1-a) and shows solitary zonal structures. As $|\psi_0|$ increases [column (b)], the nonlinear interactions between the zonal structures begin to disrupt their propagation. At even larger $|\psi_0|$ [columns (c) and (d)] the zonal structures cease to propagate quickly after they develop, in contrast to the NLSE cases [(1-c) and (1-d)]. Again, we notice that the quasilinear and nonlinear simulation results are qualitatively similar, demonstrating the sufficiency of the quasilinear approximation for our purposes.

In summary, in the nonlinear stage of the MI of coherent DWs, the NLSE does not properly approximate the (quasilinear) mHME when the primary-wave amplitude is relatively large. Only when the primary-wave amplitude is relatively small can propagating zonal structures similar to the DW--ZF solitons form.

\subsection{Zonal structures from incoherent drift waves}\label{sec:FZI}
\begin{figure*}
 \includegraphics[width=\textwidth]{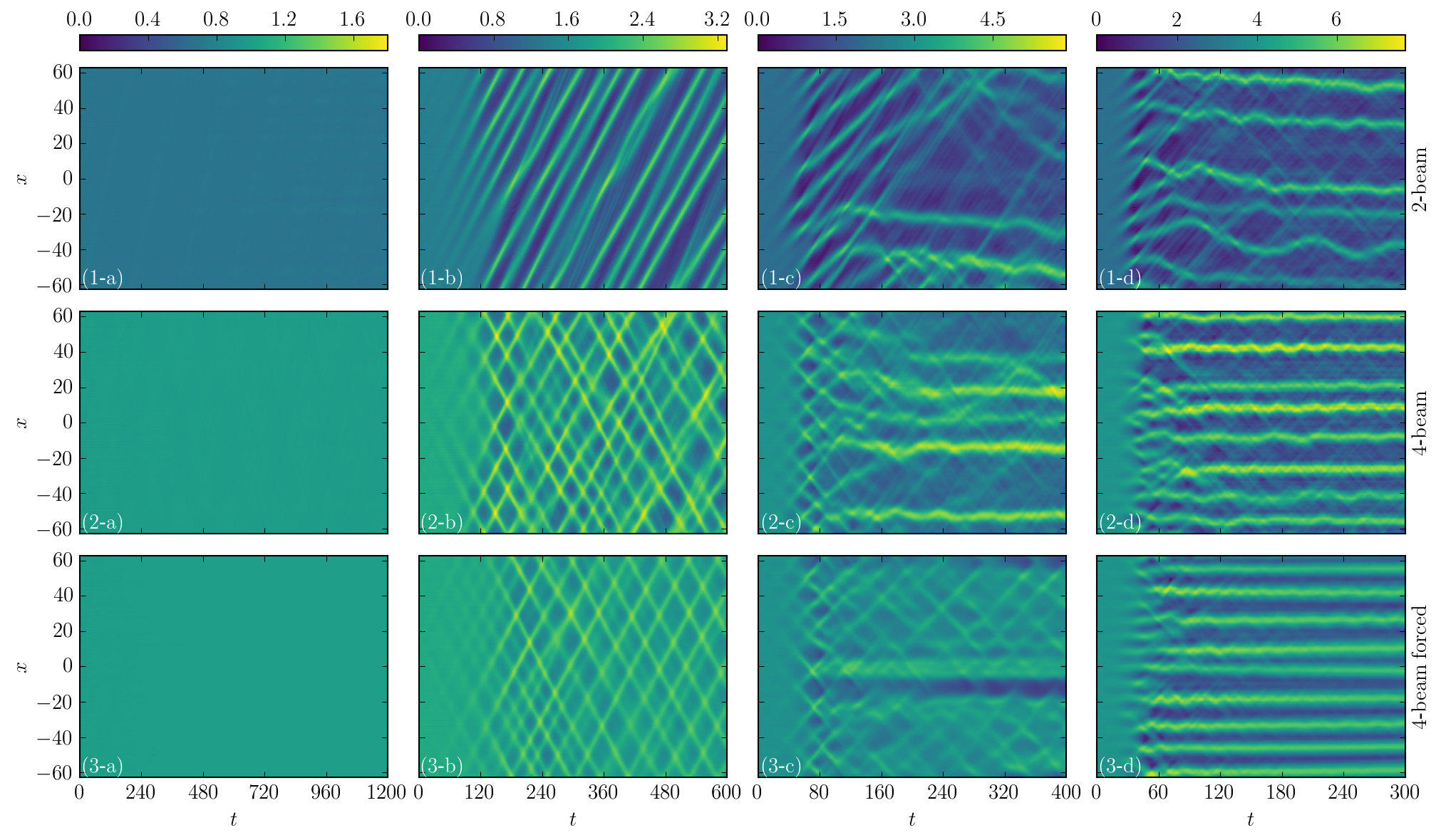}
 \caption{WM simulations of the MI of DW spectra given by Eq.\,\eqref{2beam} (row 1) and Eq.\,\eqref{4beam} (rows 2 and 3). The simulations in row 3 also include forcing and dissipation ($\mu=0.005$). In all cases, the beam width $\sigma=0.1$. The columns correspond to  various values of the effective primary-wave amplitude: (a) $|\psi_0|= 1$, (b) $|\psi_0|= 2$, (c) $|\psi_0|= 3$, and (d) $|\psi_0|= 4$. The colormaps show the spatial-temporal evolution of the DW envelope $\sqrt{\langle\tilde{w}^2\rangle}$. The simulations are run for different total time due to the differences in the growth rates. The same parameters as in Fig.\,\ref{envelope} are used. }\label{ZI}
\end{figure*}

As discussed in Sec.\,\ref{sec:ZID}, the MI of a monochromatic DW, which is studied in Sec.\,\ref{sec:FMI}, is equivalent to that of the delta-shaped DW spectrum \eqref{2delta} in the WM model. Now, let us consider a slightly more general case, where the DW spectrum has a finite width in $p_x$. Specifically, we adopt
\begin{equation}
\mathcal{W}(\mathbf{p})=\pi^2|\psi_0|^2\sum_{\pm}\delta({p}_y\pm{k}_y)f_\sigma(p_x\pm k_x),\label{2beam}
\end{equation}
with $f_\sigma(p)\doteq{\exp[{-{p}^2/(2\sigma^2)}}]/\sqrt{2\pi\sigma}$. Here, we keep the distribution in $p_y$ as delta functions for simplicity. The justification is that within the quasilinear approximation, $p_y$ is a constant of motion, and the coupling in $p_y$ is weak. For convenience, we refer to the spectrum \eqref{2beam} as ``2-beam'' for it consists of a pair of quasi-monochromatic drifton beams. (As mentioned in Sec.\,\ref{sec:WM}, driftons come in pairs.) Equation \eqref{2delta} is reproduced as the limit of Eq.\,\eqref{2beam} as the beam width $\sigma\rightarrow 0$.

In Fig.\,\ref{ZI} (row 1), we present numerical simulations of the MI of the 2-beam spectrum \eqref{2beam} with random perturbations. These simulations employ the spectral representation of the WM equation \eqref{WME} derived in the appendix of Ref.\,\cite{Ruiz2016}. The simulation domain is periodic in $x$. In contrast to Fig.\,\ref{MI}(2-a), there is no instability at small $|\psi_0|$ (1-a), which demonstrates the stabilizing effect of the finite beam width $\sigma$. Still, as the effective amplitude $|\psi_0|$ increases, the system becomes modulationally unstable, and the corresponding features are qualitatively similar to the quasilinear mHME simulations in Fig.\,\ref{MI} (row 2). With moderate $|\psi_0|$ (1-b), solitary zonal structures emerge. When $|\psi_0|$ is increased further [(1-c) and (1-d)], the zonal structures stop propagating and eventually become stationary.

As the next generalization, we consider the MI of a DW spectrum that has two pairs of quasi-monochromatic drifton beams, 
\begin{equation}
\mathcal{W}(\mathbf{p})=\pi^2|\psi_0|^2\sum_{\pm}\delta({p}_y\pm{k}_y)[f_\sigma(p_x\pm k_x)+f_\sigma(p_x\mp k_x)].\label{4beam}
\end{equation}
For convenience, we refer to the spectrum \eqref{4beam} as ``4-beam''.
In the limit as $\sigma\rightarrow 0$, this spectrum corresponds to the mixed state of two plane waves with amplitude $|\psi_0|$ and wave-vectors $(k_x,k_y)$ and $(-k_x,k_y)$, respectively. In this limit, the dispersion relation of the MI reads \cite{Ruiz2016}
\begin{align}
\prod_{\pm}&\left[\bar{k}^4\bar{k}_{\pm }^4\omega^2 -4\beta^2 q^2k_y^2(k_x\pm q/2)^2\right]\nonumber\\
=&~{|\psi_0|^2} q^2k_y^2({\bar{k}^2}-{q^2})(4k_x^2-{\bar{k}^2}-{q^2})\nonumber\\
&\times[ \bar{k}^4\bar{k}_{+}^2\bar{k}_{-}^2\omega^2+\beta^2 q^2k_y^2(4k_x^2- q^2)].
\end{align}

In Fig.\,\ref{ZI} (row 2), we show numerical simulations of the MI of the 4-beam spectrum \eqref{4beam} with random perturbations. On one hand, we observe many features common with the 2-beam case (row 1): the system is stable at small $|\psi_0|$ (2-a); solitary zonal structures form as $|\psi_0|$ increases (2-b); when $|\psi_0|$ is even larger, the zonal structures become stationary [(2-c) and (2-d)]. On the other hand, unlike in the 2-beam case, counter-propagating solitary zonal structures emerge. This is due to the fact that the group velocities of the two drifton beams have opposite signs. The implication of the 4-beam case is that solitary zonal structures can also form via the MI of multiple (pairs of) quasi-monochromatic drifton beams.

\begin{figure*}
 \includegraphics[width=\textwidth]{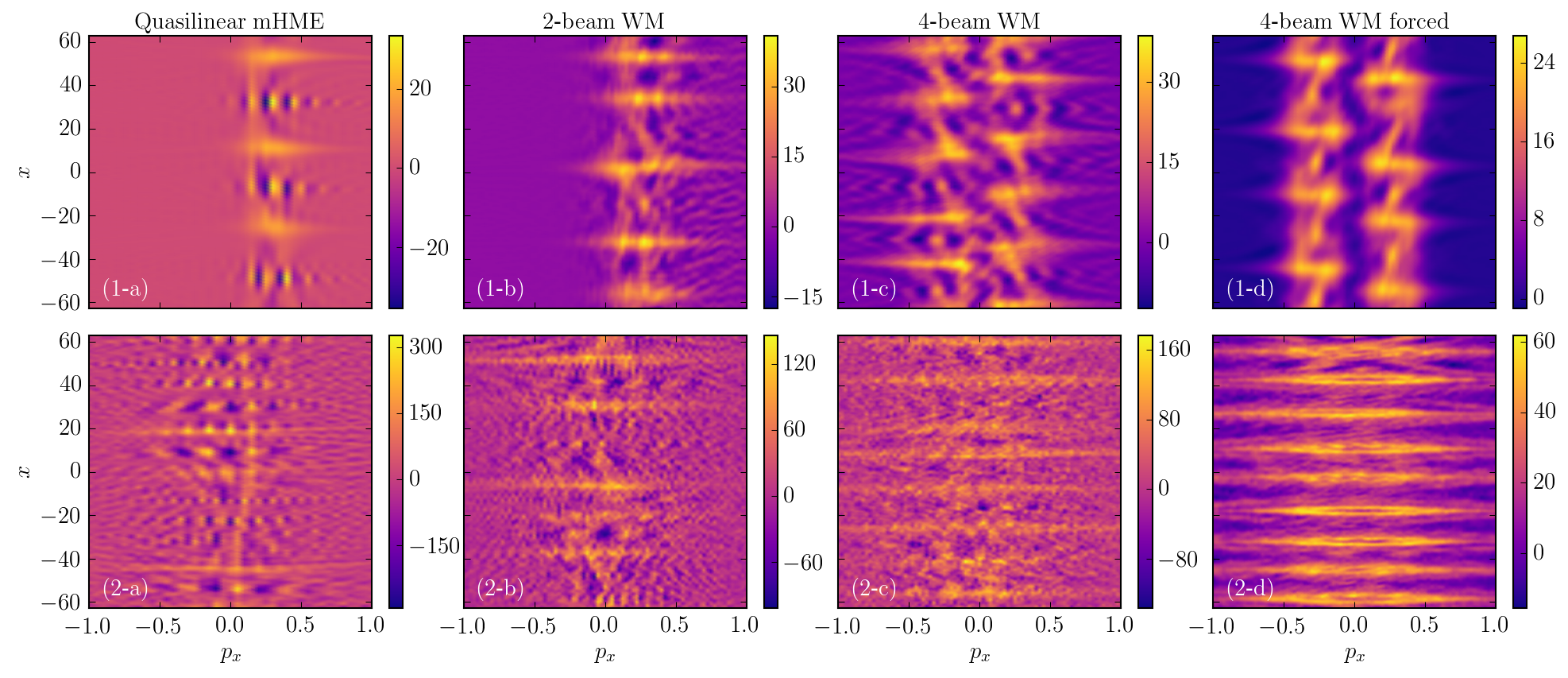}
 \caption{Snapshots of the Wigner function $W(x,p_x)|_{p_y=k_y}$ obtained from various simulations. Column (a) is from quasilinear mHME simulations, while the other columns are from WM simulations with different setups: (b)  2-beam, (c) 4-beam unforced, and (d) 4-beam forced. Row 1 is taken from the solitary zonal structures, while row 2 is from the stationary ones. Specifically, the snapshots are from: (1-a) Fig.\,\ref{MI}(2-a), $t=390$; (2-a) Fig.\,\ref{MI}(2-d), $t=300$; (1-b) Fig.\,\ref{ZI}(1-b), $t=375$; (2-b) Fig.\,\ref{ZI}(1-d), $t=300$; (1-c) Fig.\,\ref{ZI}(2-b), $t=180$; (2-c) Fig.\,\ref{ZI}(2-d), $t=300$; (1-d) Fig.\,\ref{ZI}(3-b), $t=600$; and (2-d) Fig.\,\ref{ZI}(3-d), $t=300$.
 }\label{wigner}
\end{figure*}

Furthermore, we model the 4-beam case with some external forcing $F$ added to the RHS of Eq.\,\eqref{DW}. To balance the energy input, we also add frictional damping $-2\mu W$ and $-\mu U$ to the RHS of Eqs.\,\eqref{DW} and \eqref{ZF}, respectively \cite{Ruiz2016}. In order to compare directly with the unforced 4-beam case above, we choose $F = 2\mu \mathcal{W}(\mathbf{p})$, with $\mathcal{W}$ given by the 4-beam spectrum \eqref{4beam}, such that the corresponding homogeneous equilibrium is $W=F/(2\mu)=\mathcal{W}$. The simulations are initialized with such equilibria in place and random perturbations on top, and  the results are shown in Fig.\,\ref{ZI} (row 3). Many qualitative features of the unforced case (row 2) are reproduced here: stability at small $|\psi_0|$ (3-a), solitary zonal structures at moderate $|\psi_0|$ (3-b), and stationary zonal structures at large $|\psi_0|$ (3-d). Hence, we conclude that the formation of solitary zonal structures is still possible even with forcing and dissipation, provided that the forcing spectrum consists of quasi-monochromatic peaks, producing quasi-monochromatic distributions of driftons. That being said, the forced case visibly differs from the unforced case in that the former has larger amplitudes of DWs between the zonal structures, which is maintained by the homogeneous production of driftons by the external forcing. Another distinction is that the solitary zonal structures in the forced case (3-b) are more coherent than those in the unforced case (2-b), which also owes to the fact that the external forcing tends to keep the DWs quasi-monochromatic.

\subsection{Phase-space structures}
It is instructive to examine these zonal structures in phase space by studying snapshots of the Wigner function. Figure \ref{solitonW}(1-b) illustrates that a DW--ZF soliton is a quasi-monochromatic drifton condensate that is concentrated in both space and momentum (at some nonzero $p_x$). For comparison, the phase-space snapshots in Fig.\,\ref{wigner} (row 1) are taken from solitary zonal structures in various simulations. They all reveal the presence of trains of such condensates located at $p_x\approx(\pm)k_x$. In the quasilinear mHME simulation (1-a), the striations between the drifton condensates are signatures of ``cat states'', akin to those in Fig.\,\ref{solitonW}(2-b). In the 4-beam cases [(1-c) and (1-d)], the phase-space snapshots show two trains of drifton condensates with opposing group velocities, also similar to Fig.\,\ref{solitonW}(2-b), such that the zonal structures counter-propagate. 

In contrast, the phase-space snapshots in Fig.\,\ref{wigner} (row 2) are taken from stationary zonal structures in different simulations. Accordingly, in these cases, the DW quanta are mostly localized at $p_x=0$, which is consistent with the stationarity of the zonal structures. Finally, we note that the phase-space structures are more coherent in the forced cases [column (d)] than in the unforced cases [column (c)], due to the homogeneous quasi-monochromatic external forcing applied.

\section{Summary and discussion}\label{sec:summary}
In this paper, we first consider coherent DWs, by rederiving the NLSE that approximately governs their radial envelope dynamics, and subsequently, its dispersion relation of the MI and soliton solution. Using mHME simulations, both quasilinear and nonlinear, we validate the NLSE and the DW--ZF soliton solution. Then, we demonstrate that the NLSE can adequately describe the spontaneous generation of solitary zonal structures in the mHME, which takes place in the nonlinear stage of the MI, but only when the amplitude of the primary DW is relatively small. Otherwise, stationary zonal structures are formed instead.

Next, we consider the MI of incoherent DW spectra, using the recently developed WM model. We show that DW spectra that consist of quasi-monochromatic drifton beams can also be modulationally unstable to the formation of solitary zonal structures, but only when the beam intensity is moderate. At higher intensity, the zonal structures become stationary, similar to the case with coherent DWs. Meanwhile, due to the stabilizing effect of the finite beam width, the system becomes stable to modulations at lower beam intensity.

In addition, using the WM formulation, we compare the solitary zonal structures formed via the MI with the DW--ZF solitons in phase space. This approach enables extraction of information that can be obscure in configuration space, especially when the DW spectrum has multiple quasi-monochromatic peaks. (As a data analysis tool alone, it can be straightforwardly applied to systems that are more complicated than the mHME). The phase-space distributions of DW quanta show common features of quasi-monochromatic drifton condensates, which suggests that the MI-induced solitary zonal structures are essentially the DW--ZF solitons. These structures cannot be described by wave-kinetic models that are based on the ray approximation, which neglect critical quantumlike effects such as diffraction. In contrast, the WM model retains these effects, subsumes both the NLSE and the wave kinetic models, and hence can support these solitary zonal structures.

It is worthwhile to comment on the relevance of our results to Rossby-wave turbulence in geophysics. In Ref.\,\cite{Dewar2007}, the envelope dynamics within the oHME is also discussed, and the equation of state is more complicated than our Eq.\,\eqref{EOS}; namely, it is not a local but an integral equation. Hence, the applicability of our results to the Charney equation remains to be further investigated. Meanwhile, our results are readily applicable to the barotropic vorticity equation, and the only adaption needed is to replace $\hat{\bar{p}}^2$ with $\hat{{p}}^2\doteq\hat{p}_x^2+\hat{p}_y^2$, and similarly for ${\bar{p}}^2$ and ${\bar{k}}^2$.

In the future, we plan to investigate the effect of background shear flows on the DW--ZF solitons, motivated by the radially propagating coherent structures recently identified in gyrokinetic simulations of subcritical plasmas \cite{VanWyk2016,VanWyk2017,McMillan2018}. In particular, the structures in Ref.\,\cite{McMillan2018} [Fig.\,3(b) therein] seem close to quasi-monochromatic, much similar to our Fig.\,\ref{solitonW}(1-a). The (normalized) sizes and amplitudes of those structures are also comparable to those of the DW--ZF solitons discussed in this paper. However, to properly account for subcriticality would be a challenge, and it is possible that one needs to resort to more sophisticated models with primary instabilities, such as the (modified) Hasegawa--Wakatani equations \cite{Hasegawa1983,Numata2007}. 

\acknowledgments
We thank L.\,Chen and F.\,Zonca for bringing Ref.\,\cite{Guo2009} to our attention, which led to this work. The research was supported by the U.S.~Department of Energy under Contract No.~DE-AC02-09CH11466.

\appendix
\section{Limitations of wave-kinetic models}\label{sec:WKE}
Wave-kinetic models of DW--ZF dynamics invoke the ray (geometrical-optics) approximation, i.e., assume that the DW wavelength is negligible compared with the ZF wavelength. In this case, the WM model \eqref{WME} reduces to 
\begin{subequations}\label{WKE}
\begin{align}
\partial_t W &= \{\mathcal{H},W\}+2\Gamma W,\label{iDW}\\
\partial_t U &= \partial_x\int\frac{\mathrm{d}^2p}{(2\pi)^2}\frac{p_x p_y}{\bar{p}^{4}} W ,\label{iZF}
\end{align}
\end{subequations}
with the Poisson bracket $\{A,B\}= (\partial_\mathbf{x}A)\cdot(\partial_\mathbf{p}B) - (\partial_\mathbf{p}A)\cdot(\partial_\mathbf{x}B)$. 
The Hermitian and anti-Hermitian parts of the Hamiltonian are given by, respectively \cite{Ruiz2016},
\begin{subequations}\label{iHG}
\begin{align}
\mathcal{H} &= (\beta+U'') p_y/\bar{p}^2 + p_y U ,\label{iHamiltonian}\\
\Gamma &= -U'''p_xp_y/\bar{p}^4.\label{iGamma}
\end{align}
\end{subequations}
This model was first derived as the geometrical-optics limit of the CE2 equations \cite{Parker2016}. Following Refs.\,\cite{Zhu2018,Zhu2018b}, we refer to it as the improved wave-kinetic equation (iWKE). The dispersion relation of the MI in the iWKE reads \cite{Parker2016,Zhu2018}
\begin{align}
1 =\int\frac{\mathrm{d}^2p}{(2\pi)^2}\mathcal{W}(\mathbf{p})\frac{-q^2p_y^2({\bar{p}^2}-{q^2})({\bar{p}^2}-4p_x^2)}{\left(\omega \bar{p}^4+2\beta qp_yp_x\right)^2}.\label{iZID}
\end{align}
By substituting the Wigner function \eqref{2delta} into Eq.\,\eqref{iZID}, we obtain the iWKE dispersion relation of the MI of a monochromatic DW,
\begin{align}
(\omega-qv_\text{g})^2=|\psi_0|^2{k_y}\chi q^2(1-{q^2}/{\bar{k}^2})/(4\beta).\label{WKD}
\end{align}

The iWKE dispersion relation \eqref{WKD} is plotted in Fig.\,\ref{dispersion} for comparison with the NLSE dispersion relation \eqref{NLSED} and the WM dispersion relation \eqref{MID}. 
To better understand the discrepancy between the them, let us also consider the traditional wave-kinetic equation (tWKE) \cite{Smolyakov1999}. The tWKE differs from the iWKE by further neglecting the high-order derivatives of $U$ in the Hamiltonian \eqref{iHG}, such that 
\begin{subequations}\label{tHG}
\begin{align}
\mathcal{H} &= \beta p_y/\bar{p}^2 + p_y U ,\label{tHamiltonian}\\
\Gamma &= 0.\label{tGamma}
\end{align}
\end{subequations}
Accordingly, the tWKE dispersion relation for the MI can be obtained by simply neglecting the factor ${q^2}/{\bar{k}^2}$ in Eq.\,\eqref{WKD}: 
\begin{align}
(\omega-qv_\text{g})^2=|\psi_0|^2{k_y}\chi q^2/(4\beta)\label{tWKD}
\end{align}
This dispersion relation diverges at large $q$, which is typical for the tWKE. Such divergence can give rise to unphysical grid-scale ZFs in numerical simulations, as shown in Refs.\,\cite{Parker2016,Ruiz2016}, where more detailed discussions on the differences between the tWKE, the iWKE, and the WM (CE2) model can be found.

Despite this caveat, the tWKE dispersion relation \eqref{tWKD} warrants direct comparison with the NLSE dispersion relation \eqref{NLSED}, since both models neglect the same high-oder derivatives of $U$. The difference is the $\chi^2q^4/4$ term. This term is due to diffraction, which is not included in the ray approximation that the tWKE (and iWKE) is based on, but retained in the quasi-optical approximation that the NLSE invokes. Likewise, it is the balance of diffraction and self-focusing that determines the size of the solitary zonal structures discussed in Sec.\,\ref{sec:soliton}.
In summary, because of the absence of full-wave (quantumlike) effects such as diffraction, wave-kinetic models based on the ray approximation do not properly capture the MI or the solitary zonal structures discussed in the main text.

%


\end{document}